\def\simgt{$_>\atop{^\sim}$}
\def\q0{q$_0$}
\def\h0{H$_0$}

\def\cm2{cm$^{-2}$}

\def\etal{et~al.}

\def\c2{\ion{C}{2}}
\def\o1{\ion{O}{1}}

\def\fe2{\ion{Fe}{2}}
\def\mg1{\ion{Mg}{1}}
\def\sil2{\ion{Si}{2}}
\def\si4{\ion{Si}{4}}
\def\al2{\ion{Al}{2}}
\def\n5{\ion{N}{5}}

\def\approxlt{\mathrel{\spose{\lower 3pt\hbox{$\sim$}}
        \raise 2.0pt\hbox{$<$}}}
\def\approxgt{\mathrel{\spose{\lower 3pt\hbox{$\sim$}}
        \raise 2.0pt\hbox{$>$}}}

\documentstyle[psfig]{mn}

\input{psfig.sty}

\newif\ifAMStwofonts

\title[The evolution of $ \Omega_{\rm HI}$]{The evolution of $\mathbf\Omega_{\rm HI}$ and the epoch of formation of damped Lyman-$\mathbf \alpha$ absorbers.}

\author[C. P\'eroux et al.]
        {C. P\'eroux$^{1}$, R. G. McMahon$^{1}$, L.
        J. Storrie-Lombardi$^2$ and M. J. Irwin$^{1}$
\\$1$ Institute of Astronomy, Madingley Road, Cambridge CB3 0HA, UK.
\\$2$ SIRTF Science Center,
California Institute of Technology, MS 100-22, Pasadena CA 91125,
USA.\\
email: celine, rgm, mike@ast.cam.ac.uk; lisa@ipac.caltech.edu}

\date{{\it In press}}

\pagerange{\pageref{firstpage}--\pageref{lastpage}}
\pubyear{2001}

\begin{document}

\maketitle

\label{firstpage}

\begin{abstract}
We present a study of the evolution of the column density
distribution, $f(N,z)$, and total neutral hydrogen mass in high-column
density quasar absorbers using candidates from a recent high-redshift
survey for damped Lyman-$\alpha$ (DLA) and Lyman limit system (LLS)
absorbers.  The observed number of LLS (N(HI)$>$ $1.6
\times10^{17}~$atom~cm$^{-2}$) is used to constrain $f(N,z)$ below the
classical DLA Wolfe \etal\ (1986) definition of $2\times 10^{20}$ atom
cm$^{-2}$.  The evolution of the number density of LLS is
consistent with our previous work but steeper than previously
published work of other authors.  At z=5, the number density of
Lyman-limit systems per unit redshift is $\sim$5, implying that these
systems are a major source of UV opacity in the high redshift
Universe.  The joint LLS-DLA analysis shows unambiguously that
$f(N,z)$ deviates significantly from a single power law and that a
$\Gamma$-law distribution of the form
$f(N,z)=(f_*/N_*)(N/N_*)^{-\beta} $exp$(-N/N_*)$ provides a better
description of the observations. These results are used to determine
the amount of neutral gas contained in DLAs and in systems with lower
column density.  Whilst in the redshift range 2 to 3.5, $\sim$90\% of
the neutral HI mass is in DLAs, we find that at z$>$3.5 this fraction
drops to only 55$\%$ and that the remaining 'missing' mass fraction of
the neutral gas lies in sub-DLAs with N(HI) $\rm 10^{19} - 2 \times
10^{20}~atom~cm^{-2}$.  The characteristic column density, $N_*$,
changes from $\rm 1.6 \times 10^{21}~atom~cm^{-2}$ at z$<$3.5 to $\rm
2.9 \times 10^{20}~atom~cm^{-2}$ at z$>$3.5, supporting a picture
where at z$>$3.5, we are directly observing the formation of high
column density neutral hydrogen DLA systems from lower column density
units.  Moreover since current metallicity studies of DLA systems
focus on the higher column density systems they may be giving a biased
or incomplete view of global galactic chemical evolution at z$>$3.
After correcting the observed mass in HI for the ``missing'' neutral
gas the comoving mass density now shows no evidence for a decrease
above z=2.

\end{abstract}


\section{Introduction}
One of the fundamental phenomena still poorly understood in cosmology
is the detailed process relating to the origin of structure formation
after the epoch of recombination.  The basic dilemma is, that while
the directly observable baryonic content of galaxies at the present
epoch is concentrated in stars, in the past, logically, this must have
been in the form of gas.  Therefore the only way to obtain a self
consistent and complete picture of the galaxy formation process is to
combine studies of the star light and the star formation rate with
studies of the gas content of the Universe to learn about the
underlying metal production and gas consumption rates.

Quasar absorbers provide a powerful observational means to study the
early stages of galaxy formation independently of their intrinsic
luminosity.  The hydrogen absorbers are normally divided into three
classes according to their column density: damped Lyman-$\alpha$ (DLA)
systems with N(HI) $\geq 2 \times 10^{20}$ atom cm$^{-2}$;
Lyman-limit Systems (LLS) with N(HI) $\geq 1.6
\times 10^{17}$ atom cm$^{-2}$; and the Lyman-$\alpha$ Forest with
N(HI) ranging from $\simeq 10^{12}$ to $1.6\times 10^{17}$ atom
cm$^{-2}$. The damped Lyman-$\alpha$ absorption lines (DLAs) are of
particular importance since they contain the bulk of the neutral gas
in the Universe at high redshift and are the major directly observable
baryonic component at these redshifts.

For historical reasons based on the observed HI column density
distribution in local galactic discs, Wolfe \etal\ (1986) introduced a
defining limit for DLAs of N(HI) $\geq 2 \times 10^{20}$ atom
cm$^{-2}$. This definition arises from the fact that 21 cm
observations of local spirals show the column density drops sharply
beyond this threshold (Bosma 1981). These systems exhibit strong
damping wings, however, technically any absorption system with a
doppler parameter $b<$100 km/s and N(HI) $>10^{19}$cm$^{-2}$ will
exhibit damping wings. As we will show, this 'low redshift' DLA
definition needs to be extended at high redshift to include systems
down to $\rm 10^{19}~atom~cm^{-2}$.

\begin{figure}
\psfig{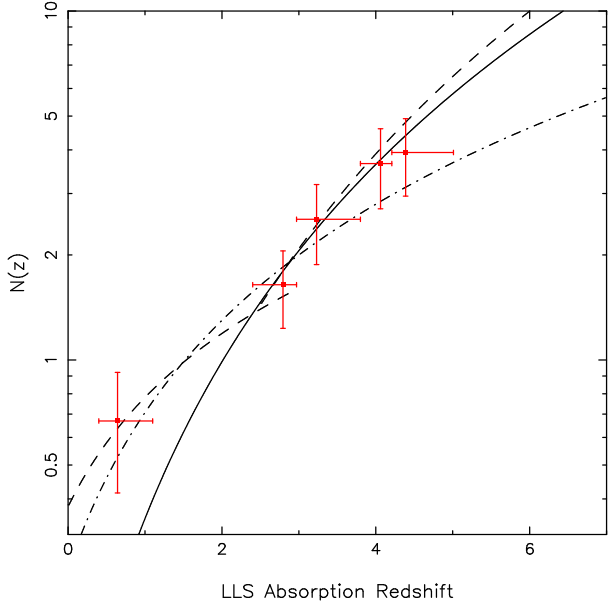}
\caption{Number of Lyman-limit systems per unit redshift. The solid
line is a fit for z $>$ 2.4.  The dashed lines are the double power
law fits from Storrie-Lombardi \etal\ (1994) and the dashed-dot line
is the fit from Stengler-Larrea \etal\ (1995). The horizontal error
bars are the bin sizes and the vertical error bars are the 1-$\sigma$
uncertainties.  The data are binned for display purpose only.}
\label{fig:LLS}
\end{figure}

\begin{figure}
\psfig{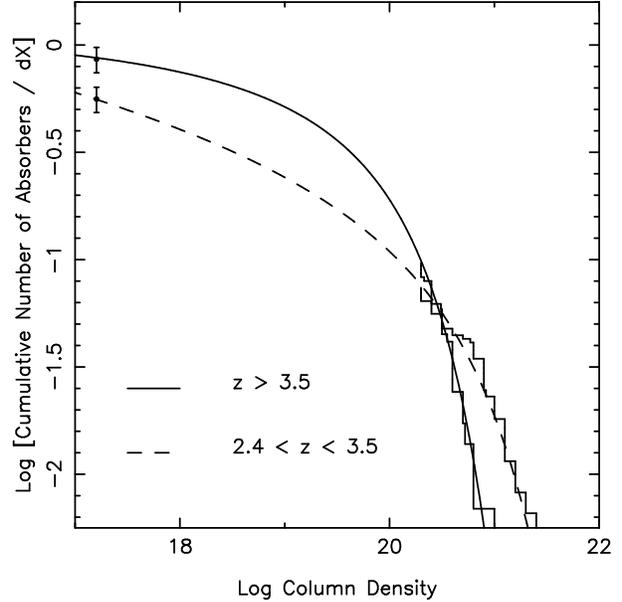}
\caption{Cumulative number of absorbers per unit distance
interval. The data point at log $N(HI) = 17.2$ atom
cm$^{-2}$ is the {\it expected number} of LLS derived from the
observed number of LLS per unit redshift. The observations are fitted
with a $\Gamma$-distribution of the form: $f(N,z) =
(f_*/N_*)(N/N_*)^{-\beta} $exp$(-N/N_*)$.}
\label{fig:cumulative}
\end{figure}

DLAs were originally thought to be the precursors of present day $L_*$
disk galaxies (Wolfe \etal\ 1986), a scenario which is supported by
some current models (Prochaska $\&$ Wolfe 1998). However, Haehnelt,
Steinmetz \& Rauch (1998) and Ledoux \etal\ (1998) have shown that the
rotating disk interpretation for DLA systems is not unique and that
the velocity structure observed by Wolfe and co-workers can also be
explained by infalling sub-galactic clumps in collapsing dark matter
halos with virial velocities of $\sim$100 $\rm km~sec^{-1}$.

LLS are a lower column density superset of DLAs, which at $z < 1$ are
probably associated with galactic halos (Steidel, Dickinson \& Persson
1994). At high redshift LLS, irrespective of their physical nature,
are an important contributor to the UV opacity of the Universe since
they essentially block all radiation shortward of 912 \AA\ in the rest
frame.

In this paper we present an analysis based on a new sample of high
column density absorbers (DLA candidates and LLS) from a recently
completed high-redshift survey (P\'eroux \etal\ 2001) combined with data
from the literature (Storrie-Lombardi \etal\ 1996c; Storrie-Lombardi
$\&$ Wolfe 2000 and refs therein).  The combined high redshift dataset
is based on observations of $\sim$100 quasars with z$>$4 and includes
complete samples of 29 DLA candidates and 37 LLS with z$>$3.5.  The
layout of this paper is as follows: \S 2 shows how the LLS can be used
to constrain the cumulative number of absorbers above log $N(HI) =
17.2$ atom cm$^{-2}$ and examines the redshift evolution of the column
density distribution.  The cosmological neutral gas evolution is
presented in \S 3 and implications of our results on theories of
structure formation are detailed in \S 4. Unless otherwise stated,
this paper assumes $\Omega_{\Lambda}=0.7$, $\Omega_{M}=0.3$ and
$H_0=65$ km s$^{-1}$ Mpc$^{-1}$.

\section{Column Density Distribution}

\begin{figure}
\psfig{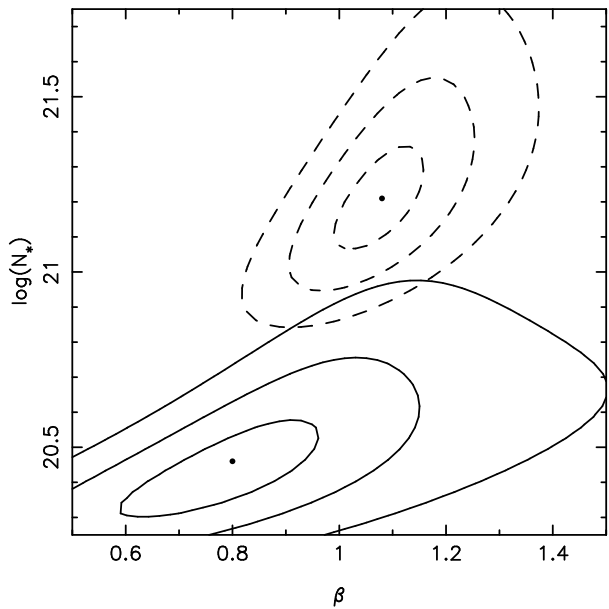}
\caption{Maximum likelihood estimator. $1$, $2$ and $3$-$\sigma$ confidence
contours to the $\Gamma$-distribution fit of the observed number of
absorbers in the redshift range $2.4 < z < 3.5$ (dashed lines) and $z >
3.5$ (solid lines).}
\label{fig:gamma-maxlike}
\end{figure}

\begin{table}
\centering
\caption{Parameter fits to the column density distribution, $f(N,z)$,
for absorbers with $\log N(HI) > 17.2$ atom cm$^{-2}$ (see equation
3). N(QSO) is the number of quasar observed and N(DLA) refers to the
number of confirmed/candidate Damped Lyman-$\alpha$ detected.}  \label{tab:1}
\begin{tabular}{cccccccc}
\hline
z&  $\beta$ & $f_*$ & $\log$ &N &N&dX &$\Omega_{DLA}$\\
Range&&$\times 10^{2}$ & $N_*$ &QSO&DLA& &$\times 10^{3}$\\
\hline
0.0-2.0       &0.74   &8.70     &20.76  &537    &23     &362.8 &0.64\\
2.0-2.7       &1.08   &3.25     &21.27  &380    &34     &522.3 &1.04\\
2.7-3.5       &1.10   &4.06     &21.18  &251    &28     &414.9 &0.98\\
\hline
2.4-3.5       &1.08   &4.29     &21.21  &314    &46     &608.2 &1.13\\  
$>$ 3.5       &0.80   &25.1     &20.46  &112    &29     &290.2 &0.71\\
\hline
\end{tabular}
\end{table}

Our new sample of high redshift Lyman-limits systems (P\'eroux \etal\
2001) is combined with data collected from previous surveys
(Storrie-Lombardi \etal\ 1994) in order to determine the number of LLS
per unit redshift (Figure~\ref{fig:LLS}).  We have chosen to use LLS
with $z\geq 2.4$ in our analysis because of the very small number of
systems known at low redshift ($0.4\leq z \leq 2.4$), all of which are
below $z=1.1$. Moreover, the paper is mainly concerned with the
evolution of the high column density absorber population at z$>$2.

In the case of the colour selected quasars observed by
Storrie-Lombardi \etal\ (1996c), P\'eroux \etal\ (2001) objects with $z
< 4.2$ are excluded to minimise colour selection bias. Only LLS with
optical depth $\tau > 1$ are used resulting in 67 systems in 124
quasars without broad absorption features, {\it not} within $3 000$ km
s$^{-1}$ of the quasar emission redshift. The latter systems are
excluded as they might be associated with the quasar itself
(Storrie-Lombardi \etal\ 1994). Details on the LLS selection process
and redshift estimates can be found in P\'eroux \etal\ (2001).  Our new
sample adds 26 LLS systems at z$>$3.5 to the 11 systems used by
Storrie-Lombardi
\etal\ (1994).  The analysis follows that of Storrie-Lombardi \etal\
(1994) and the LLS number density is fitted using a power law of the
form $N(z)=N_o(1+z)^{\gamma}$.  The parameter values for our fit,
$N_o=0.07^{+0.13}_{-0.04}$ and $\gamma=2.45^{+0.75}_{-0.65}$, are
determined using a maximum likelihood analysis.  Interestingly the
number density distribution in the Lyman-$\alpha$ forest shows a break
at $z \sim 1.5$ below which the distribution is flat ($\gamma \sim
0.2$, Weymann \etal\ 1998) and above which the distribution is steep
($\gamma=2.19 \pm 0.27$, Kim, Cristiani \& D'Odorico 2001), comparable
to the LLS results.  Our new determination agrees very well with the
previous results of Storrie-Lombardi \etal\ (1994) but there is a
significant difference at high redshift when compared with the
extrapolated results of Stengler-Larrea \etal\ (1995). This is not
surprising since their analysis contained no data above z=3.5

\begin{figure}
\psfig{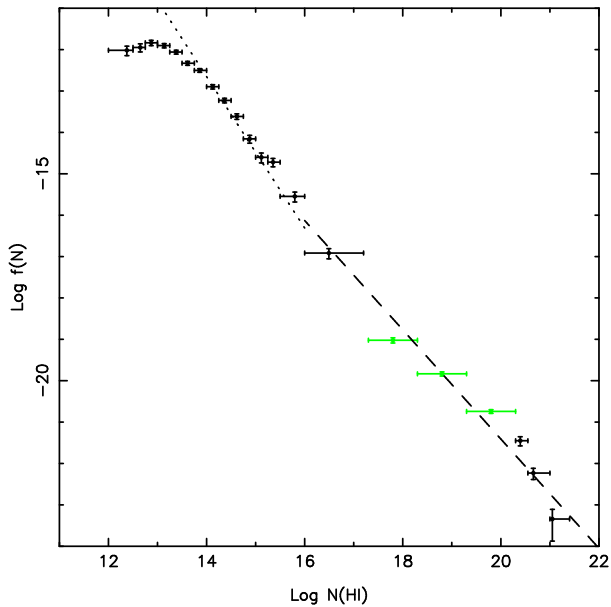}
\caption{Column density distribution, $f(N,z)$, at $z_{\rm abs} >
3.5$. The low column density data are Keck-HIRES observations of the
Lyman-$\alpha$ forest (BR 1033$-$0327 and Q0000$-$26, Williger \etal\
1994 and Lu \etal\ 1996, respectively). The light grey bins (in the
range $17.2 < $ log $ N(HI) < 20.3$) are deduced from the fit to the
observed cumulative number of quasar absorbers. The turn-over at the
low column density end is incompleteness due to a combination of
spectral resolution and signal-to-noise. The dashed and dotted lines
are the two $<z> \simeq 2.8$ power law fits from Petitjean \etal\
(1993) corrected for the absorber number density evolution with
redshift and to $\Omega_{\Lambda}=0.7$, $\Omega_{M}=0.3$ cosmology.}
\label{f:fN}
\end{figure}

We now analyse the combined samples of Lyman-limit systems and
confirmed/candidate damped Lyman-$\alpha$ systems. The cumulative
number of absorbers per unit distance interval for two redshift ranges
is presented in Figure~\ref{fig:cumulative}. The absorption distance
interval, $dX$, (Bahcall \& Peebles, 1969; Tytler, 1987) is used to
correct to co-moving coordinates and thus depends on the geometry of
the Universe since

\begin{equation}
X(z) = \int_{0}^{z} (1 + z)^2 \left[(1 + z)^2 (1 +
z\Omega_M) - z (2 + z) \Omega_{\Lambda}\right]^{-1/2}dz
\end{equation}

This equation differs from equation (3) in Storrie-Lombardi \etal\
(1996a) in that we have included the terms for a non-zero Lambda
Universe. The data with N(HI) $\geq 2 \times 10^{20}$ atom cm$^{-2}$
are DLA candidates taken from our recent high-redshift survey (our
observations more than double the redshift path surveyed at $z
$\simgt$ 3.5$ -- see P\'eroux \etal\ 2001) and DLAs published by
Storrie-Lombardi \& Wolfe 2000. The data used for the analysis are
tabulated in Appendix A \& B.

The
power law fit to the observed number of LLS per unit redshift is used
to calculate the {\it expected number} of LLS systems:

\begin{equation}
LLS_{expected}= \sum_{i=1}^{n}
\int_{z_{min}}^{z_{max}}N_o(1+z)^{\gamma}dz
\end{equation}

where $z_{min}$ and $z_{max}$ define the redshift path along which
quasar absorbers were searched for. The LLS line profiles cannot be
used to directly measure their column densities because in the range
$1.6 \times 10^{17}$ to $2 \times 10^{20}$ atom cm$^{-2}$ the curve of
growth is degenerate. Nevertheless, the {\it expected number} of LLS
provides a further constraint on the cumulative number of quasar
absorbers and clear evidence that a simple power law is {\it not} a
good representation of the observations. We thus choose to fit the
data with a $\Gamma$-distribution (cf. Schechter (1976) function used
in studies of the galaxy luminosity function) as introduced by Pei
$\&$ Fall (1995) and Storrie-Lombardi, Irwin \& McMahon (1996b):

\begin{equation}
f(N,z) = (f_*/N_*)(N/N_*)^{-\beta} e^{-N/N_*} 
\end{equation}

where $N$ is the column density, $N_*$ a characteristic column density
and $f_*$ a normalising constant. A maximum likelihood analysis is
used to derive the parameters in various redshift ranges (see
Table~\ref{tab:1}). The $1$, $2$ and $3$-$\sigma$ confidence contours
are shown in Figure~\ref{fig:gamma-maxlike} for $z < 3.5$ and $z >
3.5$. The distributions are clearly different at the $\sim$3$\sigma$
level and indicate that there are less high column density systems
(N(HI) $> 10^{21}$ atom cm$^{-2}$) at high-redshift, z$>$3.5, compared
with 2$<$z$<$3.5 confirming the results from Storrie-Lombardi, McMahon
\& Irwin (1996a) and Storrie-Lombardi \& Wolfe (2000). For comparison 
with previous works, Figure~\ref{f:fN} shows the column density
distribution, $f(N,z)$, at $z_{\rm abs}>3.5$ together with the double
power law of Petitjean et al. (1993).

\section{Cosmological Mass Density Evolution}

The mass density of absorbers can be expressed in units of the current
critical mass density, $\rho_{crit}$, as:

\begin{equation}
\Omega_{DLA}(z) = \frac{H_o \mu m_H}{c \rho_{crit}}
\int_{N_{min}}^{\infty} N f(N,z) dN
\end{equation}
where $\mu$ is the mean molecular weight and $m_H$ is the hydrogen
mass. The total HI may be estimated as:

\begin{equation}
\int_{N_{min}}^{\infty} N f(N,z) dN = \frac{\sum N_i(HI)}{\Delta X}
\end{equation}

\begin{figure}
\psfig{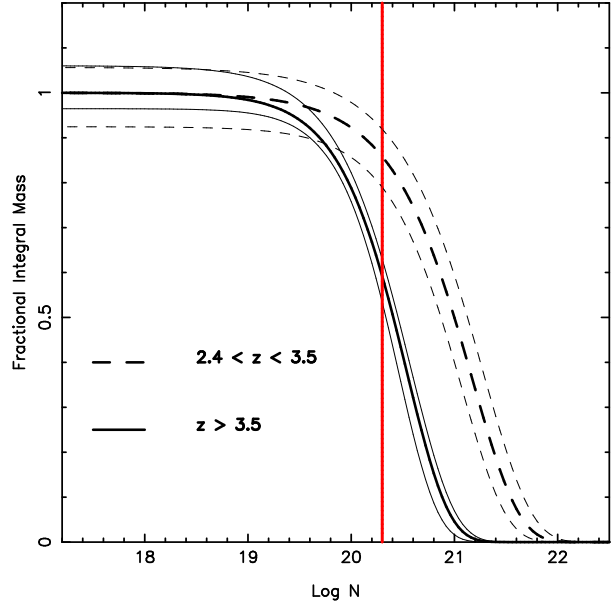}
\caption{Fractional mass integral plot for two different redshift
ranges ($z < 3.5$ and $ > 3.5$). The fine lines represent the
uncertainties in the model fit. The vertical solid line indicates the
boundary of the DLA definition. This plot shows that at $2.4 < z <
3.5$ most of the mass is contained in DLA absorbers with $N(HI) \geq 2
\times 10^{20}$ atom cm$^{-2}$, while at $z > 3.5$, $\sim 45\%$ of the
mass is under this formal limit. }
\label{fig:integral-HI}
\end{figure}

If a power law is used to fit $f(N,z)$, up to $90\%$ of the neutral
gas is in DLAs (Lanzetta, Wolfe \& Turnshek 1995), although an
artificial cut-off needs to be introduced at the high column density
end because of the divergence of the integral. If instead a
$\Gamma$-distribution is fitted to $f(N,z)$ this removes the need to
artificially truncate the high end column distribution and can be used
to probe in more detail the neutral gas fraction as a function of
column density and how this changes with redshift.
Figure~\ref{fig:integral-HI} shows the cumulative mass fraction as a
function of column density for $z >$ and $< 3.5$. At $z$ \simgt $3.5$,
up to $45\%$ of the neutral gas is in systems with $10^{19}< N(HI) <
10^{20.3}$.  We refer to these systems as {\it sub-DLAs}.  These
observations supports a picture where at z$>$3.5, we are directly
observing the formation of high column density neutral hydrogen DLA
systems from lower column density units.  The value of $N_*$ decreases
by a factor of $\approx$5 from $\rm 1.6 \times 10^{21}~atom~cm^{-2}$
at z$<$3.5 to $\rm 2.9 \times 10^{20}~atom~cm^{-2}$ at z$>$3.5.

\begin{figure}
\psfig{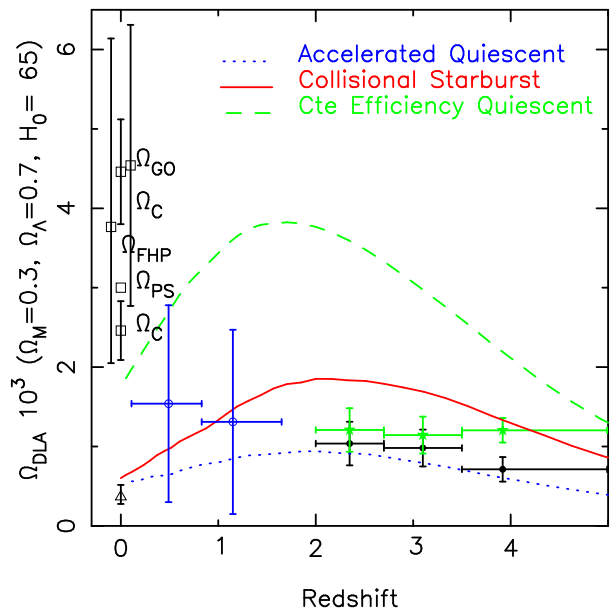}
\caption{The circles show the neutral gas in damped
Lyman-$\alpha$ galaxies in a $\Omega_{\Lambda}=0.7$, $\Omega_{M}=0.3$
and $h=0.65$ Universe. Vertical error bars correspond to 1-$\sigma$
uncertainties and the horizontal error bars indicate bin sizes. The
light grey stars are the total HI+HeII including a correction for the
neutral gas not contained in DLAs.  The open circles at low redshift
are the measurements from Rao $\&$ Turnshek (2000).  The triangle at
$z=0$ is the local HI mass measured by Natarajan $\&$ Pettini (1997)
who used the most recent galaxy luminosity function to confirm results
from Rao $\&$ Briggs (1993). The squares, $\Omega_{FHP}$,
$\Omega_{GO}$, $\Omega_{C}$, $\Omega_{PS}$ (Fukugita, Hogan $\&$
Peebles 1998, Gnedin $\&$ Ostriker 1992 and Cole \etal\ 2000, Persic
\& Salucci 1992) respectively) are $\Omega_{stellar}$ in local
galaxies. The $\Omega_{C et al.}$ error-bar plotted here does not
include uncertainties in the galactic mass-to-light
ratio. Semi-analytical models which vary in their recipe for star
formation are overplotted (Somerville, Primack $\&$ Faber 2000). These
represent the cold gas (molecular plus neutral) and thus should lie
above the observations.}
\label{fig:Omega-HI}
\end{figure}

Figure~\ref{fig:Omega-HI} displays $\Omega_{DLA}$ contained in DLAs
(filled circles) and the total amount of neutral gas (DLAs plus
sub-DLAs) for a non-zero $\Lambda$ Universe (grey stars). Our
observations are consistent with no evolution in the redshift range z=
2 to 5. Empty Universe models produce similar
results.

\section{Discussion and conclusions} 

Our study is mainly based on optically selected quasars, so it is
obvious that quasars that lie behind dusty DLA systems will be
underrepresented if they exist. Pei \& Fall (1995) have used
self-consistent closed-box/inflow-outflow galactic chemical evolution
models to show that the fraction of missing DLAs at $z=3$ ranges from
23\% to 38\%.  However, dust is unlikely to be important at
high-redshift because of the short time available for its production.
Dust will have a larger effect at z=2 compared with z$>$3 and thus it
could not cause the form of evolution in f(N,z) that we observe. But a
recent work by Ellison et al. (2001) has shown that measurements of
$\Omega_{DLA}$ based on optically selected quasars survey are not
significantly affected by dust at z$\sim$2.

It is possible that the overall cosmological evolution of
$\Omega_{DLA}$ is dominated by feedback processes rather than by
gas consumption due to star formation. The observed mass of HI at any
redshift may relate more to the recombination timescale for the
ionized HII and the timescale for cooling and collapse into molecular
hydrogen ($\rm H_2$) and thereafter into stars.  Models
from Efstathiou (2000) indicate that DLA systems might predominantly
be due to the outer parts of galaxies which do not even participate in
star formation.

It can be seen in Figure~\ref{fig:Omega-HI} that $\Omega_{DLA}$ is
significantly below the current estimates of $\Omega_M$ in stars in
the nearby Universe. This is a significant change in the situation
compared with Storrie-Lombardi \etal\ 1996c (Figure~1b). The main
reason for this change is that for the currently favoured
Lambda-dominated cosmology, the mass in HI at high redshift drops by a
factor of $\sim$50\% compared with an $\Omega_M$=1 Universe.  This is
purely a geometric effect and mainly affects the relative
normalisation between z=2 and z=0.

In any case, the observations in the redshift range 2 to 5 show no
evolution in the {\it total} amount of neutral gas in contrast to the
earlier results of Lanzetta, Wolfe and Turnshek (1995), who found that
$\Omega_{DLA}$(z$\sim$3) was twice $\Omega_{DLA}$(z$\sim$2).  Under
simple assumptions of closed box evolution, this could be interpreted
as indicating there is little gas consumption due to star formation in
DLA systems in this redshift range. Similarly, at z $>$ 2, Prochaska,
Gawiser \& Wolfe (2001) conclude that there is no evolution in the
metallicity of DLA systems from column density-weighted Fe abundance
measurements in DLAs.  Since metallicity studies focus on the higher
column density systems they may be giving a biased or incomplete view
of global galactic chemical evolution at z$>$3.  These metallicity
evolution results are still very much open to debate as another study
by Savaglio (2000) shows that the metallicity content of DLAs and
sub-DLAs {\it does} decrease with redshift when one excludes the
highest column density systems (N(HI)$>6 \times 10^{20}$) from the
analysis. It is important to note that at z$>$3.5, 90\% of the HI lies
below this limit.  Moreover, the current practice of using column
density weighted metallicities neglects the fact that the metallicity
observations are biased towards high HI column density systems and
hence do not necessarily trace the global metallicity evolution.

The detailed comparison of high redshift (z$>$2) observations with
lower redshifts is currently quite difficult. In the local Universe
21-cm emission measurements are used to determine the neutral gas mass
density. However, direct detections of DLA systems are paradoxically
more difficult for several reasons: the observed DLA wavelengths are
shifted to the UV requiring HST observations, the redshift path
available per quasar is lower and the number of DLA systems per unit
redshift is lower. In fact the most recent determination of
$\Omega_{DLA}$ at z$\sim$1 by Rao \& Turnshek (2000) is higher than
the previous estimate by Lanzetta, Wolfe and Turnshek (1991).

Nevertheless, since DLAs are the most directly observable baryonic
mass systems at high redshift, their properties are an important
constraint for theories of galaxy formation (Kauffmann $\&$ Charlot
1994, Kauffmann $\&$ Haehnelt 2000). For example, Somerville, Primack
$\&$ Faber (2000) have used the integrated total neutral gas (HI+HeII)
to constrain models (Figure~\ref{fig:Omega-HI}) which vary in their
recipes for star formation (due to collisional starburst, constant
efficiency quiescent star formation or accelerated quiescent star
formation). The models take into account the ``cold gas'' which
includes neutral as well as molecular gas.

To summarise, we find that at z$>$3.5 the fraction of neutral gas mass
in DLAs above the commonly used Wolfe \etal\ (1996) DLA definition of
N(HI)$> 2 \times 10^{20}$ atom cm$^{-2}$ is only 55$\%$ compared with
$\sim$90\% at z$<$3.5 and that the remaining fraction of the neutral
gas mass lies in systems below this limit, in ``sub-DLAs'' with column
density $ 10^{19} <$ N(HI) $< 2 \times 10^{20}$ atom cm$^{-2}$. We
also find that the total mass is conserved over the redshift range 2
to 5. This is strongly indicative that we are observing the assembly
of high column density systems from lower column density units and
independently of the precise physical nature of damped Lyman-$\alpha$
shows that we are observing the epoch of their formation or initial
collapse. A systematic study of the kinematic and metallic properties
of DLA systems with z$>$3.5 and N(HI) above $10^{19}$ atom cm$^{-2}$
is an obvious program for the new generation of echelle spectographs
on 8m class telescopes. It will also be important to directly
establish the N(HI) column density distribution function below 2
$\times 10^{20}$ atom cm$^{-2}$. We are currently undertaking such a
program using VLT UVES archival echelle data of high redshift quasars
(Dessauges-Zavadsky \etal\ 2003, P\'eroux \etal\ 2003).  The data used
for the analysis in this paper are available from
http://www.ast.cam.ac.uk/\~{}quasars.

\section{Acknowledgments}

We thank Bob Carswell, Mike Fall, Sandro D'Odorico, Patrick Petitjean,
Max Pettini, Martin Rees, and Joop Schaye for helpful discussions. CP
is grateful to PPARC and the Isaac Newton Trust for support and RGM
thanks the Royal Society and the Raymond and Beverly Sackler Foundation
for support.

\appendix

\section{The sample of quasars with confirmed or candidate DLA absorbers}

Table~\ref{tab:2} lists the 114 confirmed or candidate DLA absorbers
used in our study, their redshift and HI column densities together
with the redshift path surveyed for each quasars.  The following
corrections to the low redshift sample derived by Lanzetta, Turnshek
and Wolfe (1995) have been made: the Q1329+4117 DLA at
$z_{abs}$=0.5193, and the Q2112+059 DLA at $z_{abs}$=0.2039 have been
ruled out (Jannuzi et al. 1998, Fynbo, Moller \& Thomsen, 2001), while
Q0302$-$223 has a DLA at $z_{abs}$=0.1014 (Jannuzi et al. 1998).

\begin{table*}
\centering
\caption{Quasars with confirmed/candidate DLAs}  \label{tab:2}
\begin{tabular}{lllllll}
\hline
Quasar	&$z_{\rm em}$	&$z_{\rm min}$	&$z_{\rm max}$	&$z_{\rm abs}$	&log N(HI)	&Ref$^a$\\
\hline
Q 0000-2619       &4.11    &2.389   &4.060   &3.3901  &21.4   &5,13   \\    
BR J0006-6208     &4.455   &2.944   &4.400   &2.97    &20.7   &   25  \\
                  &        &        &        &3.20    &20.9   &25     \\
                  &        &        &        &3.78    &21.0   &25     \\
Q 0010-0012       &2.15    &1.634   &2.119   &2.0233  &20.8   &4,10   \\    
Q 0013-0029       &2.08    &1.634   &2.049   &1.9730  &20.7   &4,11   \\    
BR B0019-1522     &4.528   &2.97    &4.473   &3.4370  &20.92  &22,1   \\    
Q 0027+0103       &2.29    &1.634   &2.257   &1.9375  &20.6   &4,10   \\    
BR J0030-5129     &4.174   &2.304   &4.122   &2.45    &20.8   &	  25  \\
Q 0042-2930       &2.39    &1.591   &2.354   &1.931   &20.5   &4      \\    
Q 0049-2820       &2.26    &1.638   &2.223   &2.0713  &20.5   &4,11   \\    
Q 0056+0125       &3.16    &2.197   &3.119   &2.7750  &21.0   &4,10   \\    
Q 0058-2914       &3.07    &1.778   &3.052   &2.6711  &21.2   &21     \\    
Q 0100-3105       &2.64    &1.687   &2.605   &2.131   &20.5   &4      \\    
Q 0100+1300       &2.69    &1.64    &2.74    &2.3093  &21.4   &16,15  \\    
Q 0102-1902       &3.04    &2.044   &2.995   &2.3693  &21.0   &5,8    \\    
Q 0102-0214       &1.98    &1.649   &1.949   &1.7431  &20.6   &4,10   \\    
PSS J0106+2601    &4.309   &2.764   &4.256   &3.96    &20.5   &	  25  \\
BRI B0111-2819    &4.30    &2.709   &4.247   &3.1043  &21.0   &1      \\    
Q 0112-3041       &2.99    &1.881   &2.945   &2.4191  &20.5   &21,9   \\    
Q 0112-3041       &2.99    &1.881   &2.945   &2.7023  &20.3   &21,9   \\
Q 0112+0300       &2.81    &1.813   &2.785   &2.4227  &21.0   &5,11   \\    
PSS J0132+1341    &4.147   &2.844   &4.096   &3.93    &20.3   &1      \\    
PSS J0133+0400    &4.154   &2.865   &4.102   &3.69    &20.4   &	  25  \\
                  &        &        &        &3.77    &20.5   &	  25  \\
PSS J0134+3307    &4.532   &2.562   &4.477   &3.76    &20.6   &	  25  \\
Q 0149+3335       &2.43    &1.64    &2.43    &2.1413  &20.5   &16,7   \\    
PSS J0152+0735    &4.051   &1.890   &4.000   &3.84    &20.7   &	  25  \\
Q 0201+3634       &2.49    &1.632   &2.879   &2.4614  &20.4   &21,8   \\    
Q 0201+3634       &2.49    &1.632   &2.879   &1.768   &20.5   &21     \\
PSS J0209+0517    &4.174   &2.759   &4.122   &3.66    &20.3   &	  25  \\
                  &        &        &        &3.86    &20.6   &	  25  \\
Q 0216+0803       &3.00    &1.731   &2.953   &2.2930  &20.5   &5,9    \\    
BR J0301-5537     &4.133   &2.825   &4.082   &3.22    &20.3   &	  25  \\
Q 0302-223        &1.4000  &1.0077  &1.3760  &1.0104  &20.36  &24     \\
BR J0307-4945     &4.728   &3.138   &4.671   &4.46    &20.8   &	  25  \\
SDSS J0310-0014   &4.658   &3.087   &4.601   &3.42    &20.5   &	  25  \\
BR B0331-1622     &4.38    &2.868   &4.326   &3.56    &20.6   &1      \\    
BR J0334-1612     &4.363   &3.080   &4.309   &3.56    &21.0   &	  25  \\
Q 0336-0142       &3.20    &2.109   &3.155   &3.0619  &21.2   &21,8   \\    
SDSS J0338+0021   &5.010   &3.528   &4.950   &4.06    &20.4   &	  25  \\
Q 0347-3819       &3.23    &2.044   &3.186   &3.0244  &20.8   &21,18  \\    
BR J0426-2202     &4.320   &2.544   &4.267   &2.98    &21.1   &	  25  \\
Q 0449-1330       &3.097   &2.006   &3.056   &2.052   &20.4   &21     \\    
Q 0458-0203       &2.29    &1.96    &2.29    &2.0399  &21.7   &16,7   \\    
Q 0528-2505       &2.779   &1.961   &2.741   &2.1404  &21.0   &5      \\    
PSS J0747+4434    &4.430   &2.764   &4.376   &3.76    &20.3   &	  25  \\
                  &        &        &        &4.02    &20.6   &	  25  \\
Q 0834-2006       &2.75    &1.632   &2.704   &1.715   &20.4   &21     \\    
Q 0836+1122       &2.70    &1.74    &2.67    &2.4660  &20.6   &16,6   \\    
Q 0913+0715       &2.78    &1.866   &2.739   &2.6187  &20.3   &21,8   \\    
MG  0930+2858     &3.41    &2.173   &3.366   &3.24    &20.5   &2      \\    
Q 0935+4143       &1.9800  &1.0626  &1.550   &1.369   &20.3   &3      \\    
BR B0951-0450     &4.369   &2.93    &4.315   &3.8580  &20.6   &22,1   \\    
BR B0951-0450     &4.369   &2.93    &4.315   &4.2028  &20.4   &22,1   \\
BRI B0952-0115    &4.426   &2.99    &4.372   &4.0238  &20.55  &22,1   \\    
PC 0953+4749      &4.457   &3.010   &4.004   &3.403   &20.9   &1      \\    
PC 0953+4749      &4.457   &3.010   &4.004   &3.890   &21.1   &1      \\
BRI B1013+0035    &4.405   &2.61    &4.351   &3.1031  &21.1   &22,1   \\    
Q 1032+0414       &3.39    &2.067   &3.347   &2.839   &20.3   &21     \\    
PSS J1057+4555    &4.116   &2.652   &4.065   &3.05    &20.3   &1,25   \\       
BRI B1108-0747    &3.922   &2.64    &3.873   &3.607   &20.33  &22,23  \\    
BRI B1114-0822    &4.495   &3.19    &4.440   &4.2576  &20.3   &22,1   \\    
Q 1151+0651       &2.76    &1.65    &2.76    &1.7737  &21.3   &16,6   \\    
\hline
\end{tabular}
\vspace{0.5cm}
\end{table*}

\begin{table*}
\centering
\setcounter{table}{0}
\caption{{\it continued}}  
\begin{tabular}{lllllll}
\hline
Quasar	&$z_{\rm em}$	&$z_{\rm min}$	&$z_{\rm max}$	&$z_{\rm abs}$	&log N(HI)	&Ref$^a$\\
\hline
Q 1159+0132       &3.27    &1.988   &3.226   &2.6846  &21.1   &21,8   \\    
PSS J1159+1337    &4.073   &2.563   &4.022   &3.72    &20.3   &	  25  \\
BR B1202-0725     &4.694   &3.16    &4.637   &4.383   &20.49  &22,23  \\    
Q 1205+0918       &2.08    &1.634   &2.046   &1.673   &20.6   &4      \\    
Q 1209+0919       &3.30    &2.175   &3.254   &2.5835  &21.4   &21,8   \\    
Q 1210+1731       &2.54    &1.634   &2.502   &1.8920  &20.6   &4,10   \\    
Q 1215+3322       &2.61    &1.64    &2.60    &1.9989  &21.0   &16,7   \\    
Q 1223+1753       &2.92    &1.945   &2.879   &2.4658  &21.5   &4,11   \\    
Q 1232+0815       &2.57    &1.789   &2.534   &2.3376  &20.9   &4,10   \\    
Q 1240+1516       &2.28    &1.634   &2.247   &1.738   &20.7   &4      \\    
Q 1244+3443       &2.48    &1.64    &2.50    &1.8593  &20.5   &16,7   \\    
Q 1246-0217       &2.11    &1.634   &2.075   &1.779   &21.2   &4      \\    
PSS J1253-0228    &4.007   &2.498   &3.957   &2.78    &21.4   &	  25  \\
Q 1308+0105       &2.80    &1.634   &2.763   &1.762   &20.6   &4      \\    
GB 1320+3927      &2.98    &1.968   &2.940   &2.11    &20.4   &2      \\    
Q 1337+1121       &2.92    &1.86    &2.92    &2.7957  &20.9   &16,6   \\    
BRI B1346-0322    &3.992   &2.65    &3.942   &3.7343  &20.72  &22,1   \\    
Q 1347+1116       &2.70    &1.92    &2.71    &2.4709  &20.3   &16,6   \\    
Q 1409+0930       &2.86    &1.979   &2.800   &2.4561  &20.5   &21,8   \\    
PSS J1443+2724    &4.407   &2.950   &4.353   &4.216   &20.8   &1      \\    
Q 1451+1223       &3.26    &2.158   &3.207   &2.478   &20.4   &16,21  \\    
BR BI1500+0824    &3.943   &2.39    &3.894   &2.7968  &20.8   &22,1   \\    
GB 1610+2806      &3.54    &2.021   &3.498   &2.59    &20.6   &2      \\    
MG 1614+0506      &3.21    &1.984   &3.168   &2.52    &20.4   &2      \\    
PSS J1618+4125    &4.213   &2.820   &4.161   &3.92    &20.5   &	  25  \\
RX J1759+6638     &4.320   &2.804   &4.267   &3.40    &20.4   &	  25  \\
GB 1759+7539      &3.05    &1.955   &3.010   &2.624   &20.77  &2      \\    
PSS J1802+5616    &4.158   &2.891   &4.106   &3.76    &20.4   &	  25  \\
PC 2047+0123      &3.799   &2.620   &3.751   &2.7299  &20.4   &1      \\    
PSS J2122-0014    &4.114   &2.350   &4.063   &3.20    &20.3   &	  25  \\
Q 2132-4321       &2.42    &1.595   &2.386   &1.916   &20.7   &4      \\    
Q 2138-4427       &3.17    &2.107   &3.128   &2.851   &20.9   &4,12   \\    
PSS J2154+0335    &4.363   &2.979   &4.309   &3.61    &20.4   &	  25  \\
PSS J2155+1358    &4.256   &2.940   &4.203   &3.32    &21.1   &	  25  \\
Q 2206-1958       &2.56    &1.85    &2.58    &1.9205  &20.5   &16,14  \\    
Q 2206-1958       &2.56    &1.85    &2.58    &2.0763  &20.7   &16,17  \\
Q 2223-0512       &1.4040  &0.4159  &0.6310  &0.4925  &20.9   &3      \\    
Q 2223-0512       &1.4040  &0.9259  &1.3800  &        &       &3      \\
Q 2230+0232       &2.15    &1.634   &2.119   &1.8642  &20.8   &4,10,11\\    
Q 2231-0015       &3.015   &1.749   &2.980   &2.0657  &20.6   &4,9    \\    
BR B2237-0607     &4.558   &2.96    &4.502   &4.0691  &20.5   &22,1   \\    
Q 2239-3836       &3.55    &2.389   &3.508   &3.2810  &20.8   &21,9   \\    
PSS J2241+1352    &4.441   &3.027   &4.387   &4.28    &20.7   &	  25  \\
Q 2248+0127       &2.56    &1.634   &2.524   &1.9080  &20.6   &4,10   \\    
BR J2317-4345     &3.943   &2.448   &3.894   &3.49    &20.9   &	  25  \\
PSS J2344+0342    &4.239   &2.696   &4.187   &3.21    &20.9   &	  25  \\
Q 2348-0108       &3.01    &2.044   &2.965   &2.4272  &20.5   &16,6   \\    
Q 2348-0108       &3.01    &2.044   &2.965   &2.6161  &21.3   &21,6   \\
Q 2351+0217       &2.03    &1.634   &2.000   &1.766   &20.9   &4,10   \\    
Q 2359-0216       &2.31    &1.747   &2.779   &2.0951  &20.7   &16,7   \\    
Q 2359-0216       &2.31    &1.747   &2.779   &2.1537  &20.3   &16,7   \\
\hline
\end{tabular}
\vspace{0.5cm}

\begin{small}

$^a$References: 1) Storrie-Lombardi \& Wolfe 2000; 2) Storrie-Lombardi \& Hook 2000; 3)
   Lanzetta, Wolfe, \& Turnshek 1995; 4) Wolfe et al. 1995; 5) Sargent,
   Steidel, \& Boksenberg 1989; 6) Turnshek et al. 1989; 7) Wolfe et
   al. 1993; 8) Lu et al. 1993; 9) Lu \& Wolfe; 10) Virgilio et
   al. 1995; 11) Pettini et al. 1994; 12) Francis \& Hewett 1993; 13)
   Savaglio et al. 1994; 14) Sargent, Boksenberg, \& Steidel 1988; 15)
   Black, Chaffee, \& Foltz 1987; 16) Wolfe et al. 1986; 17) Wolfe et
   al. 1994; 18) Rauch et al. 1990; 19) Williger et al. 1989; 20)
   Meyer, Lanzetta \& Wolfe 1995; 21) Lanzetta et al. 1991; 22)
   Storrie-Lombardi, McMahon, Irwin, \& Hazard 1996; 23)
   Storrie-Lombardi, Irwin \& McMahon 1996; 24) Jannuzi et al. 1998; 25)
   P\'eroux et al. 2001
\end{small}
\end{table*}

%%%%%%%%%%%%%%%%%%%%%%%%%%%%%%%%%%%%%%%%%%%%%%%%%%%%%%%%%%%%%%%%%%%%%%%%%%%%%%%%%5
%%%%%%%%%%%%%%%%%%%%%%%%%%%%%%%%%%%%%%%%%%%%%%%%%%%%%%%%%%%%%%%%%%%%%%%%%%%%%%%%%5
%%%%%%%%%%%%%%%%%%%%%%%%%%%%%%%%%%%%%%%%%%%%%%%%%%%%%%%%%%%%%%%%%%%%%%%%%%%%%%%%%5
%%%%%%%%%%%%%%%%%%%%%%%%%%%%%%%%%%%%%%%%%%%%%%%%%%%%%%%%%%%%%%%%%%%%%%%%%%%%%%%%%5

\section{The sample of quasars without DLA absorbers}

Table~\ref{tab:3} lists the quasars from our study where no DLAs where
discovered. The table also provides the redshift path surveyed for
each quasars.

\begin{table}
\centering
\caption{Quasars without DLAs detected.}  
\label{tab:3}
\begin{tabular}{lllll}
\hline
Quasar	&$z_{\rm em}$	&$z_{\rm min}$	&$z_{\rm max}$	&Ref$^a$\\
\hline
Q 0001+0842       &3.241   &2.024   &3.229   &21   	\\ 
Q 0002+151        &1.8990  &0.4723  &0.6034  &3    	\\ 
Q 0002+151        &1.8990  &1.1198  &1.5500  &3    	\\ 
Q 0003+158        &0.4500  &0.0080  &0.4355  &3    	\\ 
PSS J0003+2730    &4.240   &2.718   &4.188   &26	\\
MG 0004+1359      &3.25    &1.899   &3.207   &2    	\\ 
Q 0004+1711       &2.898   &2.002   &2.851   &21   	\\  
Q 0006+0230       &2.09    &1.787   &2.059   &4    	\\
Q 0006+0200       &2.35    &1.634   &2.317   &4    	\\
Q 0007-0004       &2.26    &1.634   &2.227   &4    	\\
Q 0007-000        &2.29    &1.670   &2.260   &16   	\\ 
MG 0007+0141      &2.90    &1.882   &2.861   &2    	\\ 
Q 0007+106        &0.0890  &0.0080  &0.0781  &3    	\\ 
Q 0009-0138       &1.99    &1.634   &1.960   &4    	\\
Q 0009+0219       &2.66    &1.784   &2.623   &4    	\\
Q 0009-0215       &2.11    &1.634   &2.079   &4    	\\
Q 0014+8118       &3.380   &1.928   &3.340   &21   	\\ 
Q 0014-0256       &1.85    &1.729   &1.821   &4    	\\
Q 0015+0239       &2.47    &1.784   &2.435   &4    	\\
Q 0016+0045       &2.31    &1.651   &2.277   &4    	\\
Q 0018-0220       &2.56    &1.634   &2.524   &4    	\\
Q 0018+0047       &1.83    &1.655   &1.802   &4    	\\
Q 0020+0217       &1.80    &1.665   &1.772   &4    	\\
Q 0022+0150       &2.77    &1.791   &2.732   &4    	\\
Q 0023+0010       &1.90    &1.657   &1.871   &4    	\\
Q 0025-0151       &2.08    &1.634   &2.049   &4    	\\
Q 0026+0158       &1.89    &1.727   &1.861   &4    	\\
Q 0026+129        &0.1420  &0.0080  &0.1306  &3    	\\ 
Q 0027+0149       &2.33    &1.694   &2.297   &4    	\\
Q 0028+0236       &2.00    &1.634   &1.970   &4    	\\
Q 0028-0148       &2.08    &1.840   &2.049   &4    	\\
Q 0029+0017       &2.23    &1.725   &2.198   &4    	\\
Q 0029-0152       &2.39    &2.013   &2.356   &4    	\\
PSS J0030+1702    &4.282   &2.763   &4.229   &1    	\\ 
PSS J0034+1639    &4.293   &2.981   &4.240   &26	\\
SDSS J0035+0040   &4.747   &3.309   &4.690   &26	\\
Q 0037-018        &2.34    &1.654   &2.303   &16   	\\ 
Q 0039-2630       &1.81    &1.634   &1.782   &4    	\\
Q 0040-2917       &2.09    &1.634   &2.056   &4    	\\
Q 0041-2638       &3.045   &1.657   &3.029   &21   	\\ 
Q 0041-2707       &2.79    &1.668   &2.748   &4    	\\
Q 0041-2607       &2.79    &1.634   &2.470   &4    	\\
Q 0041-2658       &2.46    &1.634   &2.422   &4    	\\
Q 0041-2859       &2.13    &1.589   &2.103   &4    	\\
Q 0042-3053       &1.97    &1.634   &1.944   &4    	\\
Q 0042-2627       &3.298   &2.113   &3.253   &21   	\\ 
Q 0042-2656       &3.33    &2.215   &3.314   &21   	\\ 
Q 0042-2657       &2.90    &2.226   &2.859   &4    	\\
Q 0043-2937       &2.23    &1.656   &2.198   &4    	\\
Q 0044+030        &0.6240  &0.3386  &0.6078  &3    	\\ 
Q 0045-3002       &2.02    &1.603   &1.991   &4    	\\
Q 0045-0341       &3.138   &1.961   &3.094   &21   	\\ 
Q 0045-013        &2.53    &1.784   &2.493   &16   	\\ 
Q 0046-293        &4.014   &2.882   &3.964   &1    	\\
BRI B0046-2458    &4.15    &2.575   &4.099   &1    	\\ 
Q 0047-2759       &2.13    &1.649   &2.099   &4    	\\
Q 0047-3050       &2.97    &1.930   &2.933   &4    	\\
Q 0047-2538       &1.97    &1.591   &1.939   &4    	\\
Q 0047-2326       &3.422   &2.291   &3.378   &21   	\\ 
Q 0048-0119       &1.88    &1.634   &1.849   &4    	\\
Q 0048-2545       &2.08    &1.634   &2.051   &4    	\\
Q 0049-0104       &2.10    &1.715   &2.065   &4    	\\
Q 0049-0012       &1.95    &1.634   &1.916   &4    	\\
Q 0049+007        &2.27    &1.644   &2.238   &16   	\\ 
\hline
\end{tabular}
\end{table}

\begin{table}
\centering
\setcounter{table}{0}
\caption{{\it continued}}  
\begin{tabular}{lllll}
\hline
Quasar	&$z_{\rm em}$	&$z_{\rm min}$	&$z_{\rm max}$	&Ref$^a$\\
\hline
Q 0049+014        &2.31    &1.681   &2.276   &16   	\\ 
Q 0049+171        &0.0640  &0.0080  &0.0534  &3    	\\ 
Q 0050+124        &0.0611  &0.0080  &0.0505  &3    	\\ 
Q 0050-2523       &2.16    &1.592   &2.127   &4    	\\
Q 0051-0226       &2.53    &1.634   &2.491   &4    	\\
Q 0052-0058       &2.21    &1.634   &2.180   &4    	\\
Q 0052+251        &0.1550  &0.0080  &0.1435  &3    	\\ 
Q 0053-0134       &2.06    &1.634   &2.031   &4    	\\
Q 0053-2824       &3.616   &2.454   &3.576   &21   	\\ 
Q 0054+0200       &1.87    &1.634   &1.844   &4    	\\
Q 0054+144        &0.1710  &0.0080  &0.1593  &3    	\\ 
Q 0054-006        &2.76    &1.854   &2.724   &16   	\\ 
Q 0055+0141       &2.23    &1.651   &2.200   &4    	\\
Q 0055-2744       &2.20    &1.567   &2.163   &4    	\\
Q 0055-2629       &3.6560  &1.920   &3.609   &21   	\\ 
Q 0055-0200       &1.98    &1.782   &1.953   &4    	\\
Q 0055+0025       &1.91    &1.634   &1.885   &4    	\\
Q 0056-0241       &2.23    &1.779   &2.194   &4    	\\
Q 0057-0225       &2.01    &1.715   &1.979   &4    	\\
Q 0057-274        &3.52    &2.603   &3.475   &1    	\\
Q 0058-2604       &2.47    &1.606   &2.437   &4    	\\
Q 0058-0227       &2.23    &1.712   &2.194   &4    	\\
Q 0058+0155       &1.95    &1.634   &1.924   &4    	\\
Q 0059-0207       &2.29    &1.653   &2.257   &4    	\\
Q 0059-2625       &2.10    &1.614   &2.069   &4    	\\
Q 0059+0035       &2.55    &1.673   &2.510   &4    	\\
PSS J0059+0003    &4.16    &2.750   &4.108   &1    	\\
Q 0100+0146       &1.91    &1.692   &1.880   &4    	\\
Q 0101-2548       &1.97    &1.596   &1.943   &4    	\\
Q 0101-3025       &4.073   &1.937   &3.116   &21   	\\ 
Q 0102-0240       &1.84    &1.731   &1.818   &4    	\\
BRI B0103+0032    &4.437   &2.87    &4.383   &22   	\\ 
Q 0103-0141       &2.21    &1.634   &2.174   &4    	\\
Q 0103-2901       &2.87    &1.922   &2.831   &4    	\\
Q 0104+0030       &1.87    &1.667   &1.845   &4    	\\
PC 0104+0215      &4.171   &2.881   &4.119   &1    	\\
Q 0105-2649       &2.46    &1.667   &2.428   &4    	\\
Q 0106-0230       &2.28    &1.634   &2.246   &4    	\\
Q 0106+0119       &2.10    &1.871   &2.068   &4    	\\
Q 0107+0022       &1.97    &1.634   &1.938   &4    	\\
Q 0108+0028       &2.01    &1.733   &1.975   &4    	\\
Q 0109+022        &2.35    &1.734   &2.317   &16   	\\ 
Q 0110-0107       &1.89    &1.643   &1.860   &4    	\\
Q 0112-2728       &2.894   &1.784   &2.855   &21   	\\  
Q 0114-0856       &3.163   &1.838   &3.118   &21   	\\ 
Q 0115-3002       &3.249   &1.733   &3.207   &21   	\\ 
PSS J0117+1552    &4.244   &2.646   &4.192   &1    	\\ 
Q 0117+213        &1.4930  &0.9989  &1.4681  &3    	\\ 
Q 0119-286        &0.1170  &0.0080  &0.1058  &3    	\\ 
Q 0119-013        &0.0540  &0.0080  &0.0435  &3    	\\ 
Q 0123+257        &2.37    &1.644   &2.338   &16   	\\ 
PC 0131+0120      &3.792   &3.116   &3.744   &1    	\\
PSS J0131+0633    &4.417   &3.014   &4.363   &26	\\
Q 0132-1947       &3.130   &1.714   &3.089   &21   	\\ 
Q 0134+329        &0.3670  &0.0080  &0.1050  &3    	\\ 
BRI B0135-4239    &3.97    &2.575   &3.920   &1    	\\ 
Q 0136+010        &2.35    &1.749   &2.317   &16   	\\ 
Q 0136+1737       &2.73    &1.632   &2.679   &21   	\\ 
Q 0143-0135       &3.141   &1.673   &3.097   &21   	\\ 
GB 0148+2502      &3.10    &1.825   &3.059   &2    	\\ 
Q 0148-0946       &2.850   &1.797   &2.810   &21   	\\ 
BRI B0151-0025    &4.194   &2.74    &4.142   &22   	\\ 
Q 0153+0430       &2.993   &1.673   &2.951   &21   	\\ 
Q 0157+001        &0.1631  &0.0080  &0.1515  &3    	\\ 
\hline
\end{tabular}
\end{table}

\begin{table}
\centering
\setcounter{table}{0}
\caption{{\it continued}}  
\begin{tabular}{lllll}
\hline
Quasar	&$z_{\rm em}$	&$z_{\rm min}$	&$z_{\rm max}$	&Ref$^a$\\
\hline
Q 0159+036        &2.47    &1.644   &2.436   &16   	\\ 
Q 0205+024        &0.1564  &0.0080  &0.1448  &3    	\\ 
Q 0207-0019       &2.853   &1.756   &2.817   &21   	\\ 
SDSS J0211-0009   &4.874   &3.402   &4.815   &26	\\
Q 0215+015        &1.7150  &0.9996  &1.5500  &3    	\\ 
Q 0219+428        &0.4440  &0.0080  &0.4296  &3    	\\ 
GB 0229+1309      &2.07    &1.767   &2.039   &2    	\\ 
Q 0232-042        &1.4360  &0.0080  &0.6320  &3    	\\ 
Q 0232-042        &1.4360  &0.8733  &1.4116  &3    	\\ 
BR J0234-1806     &4.301   &2.971   &4.248   &26	\\
Q 0237-233        &2.2230  &1.1593  &1.5402  &3    	\\ 
Q 0239-1527       &2.786   &1.928   &2.744   &21   	\\ 
BRI B0241-0146    &4.053   &2.86    &4.002   &22   	\\ 
BR B0245-0608     &4.238   &2.96    &4.186   &22   	\\ 
PSS J0248+1802    &4.422   &2.810   &4.368   &1,26	\\
Q 0249-1826       &3.210   &1.871   &3.163   &21   	\\ 
Q 0249-2212       &3.21    &2.044   &3.160   &5,1  	\\ 
Q 0252+0136       &2.47    &1.634   &2.430   &4    	\\
Q 0254+0000       &2.25    &1.634   &2.215   &4    	\\
Q 0256-0000       &3.377   &2.241   &3.330   &21   	\\ 
Q 0256-0031       &2.00    &1.634   &1.965   &4    	\\
Q 0258+0210       &2.52    &1.634   &2.489   &4    	\\
Q 0301-0035       &3.226   &2.060   &3.181   &21   	\\ 
Q 0302-0019       &3.290   &1.739   &3.243   &21   	\\ 
Q 0305+0127       &2.15    &1.634   &2.118   &4    	\\
Q 0307-0058       &2.11    &1.634   &2.075   &4    	\\
Q 0308+0129       &2.34    &1.739   &2.302   &4    	\\
Q 0308+1902       &2.839   &1.673   &2.797   &21   	\\ 
Q 0308-1920       &2.756   &1.673   &2.714   &21   	\\  
BR J0311-1722     &4.039   &2.591   &3.989   &26	\\
Q 0312-770        &0.2230  &0.0080  &0.2108  &3    	\\ 
Q 0316-2023       &2.869   &1.747   &2.826   &21   	\\ 
Q 0323+022        &0.1470  &0.0080  &0.1295  &3    	\\ 
BR J0324-2918     &4.622   &2.900   &4.566   &26	\\
Q 0329-2534       &2.689   &1.661   &2.662   &21   	\\ 
Q 0334-2029       &3.132   &2.057   &3.089   &21   	\\ 
PC 0345+0130      &3.638   &2.699   &3.592   &1    	\\
BR B0351-1034     &4.351   &3.09    &4.297   &22   	\\ 
Q 0351-3904       &3.01    &1.632   &2.970   &21   	\\ 
Q 0352-2732       &2.823   &1.673   &2.781   &21   	\\ 
BR J0355-3811     &4.545   &3.030   &4.490   &26	\\
BR B0401-1711     &4.236   &2.82    &4.184   &22   	\\ 
BR J0403-1703     &4.227   &2.992   &4.175   &26	\\
Q 0405-123        &0.5740  &0.0080  &0.5583  &3    	\\ 
Q 0414-060        &0.7810  &0.0080  &0.7632  &3    	\\ 
BR J0415-4357     &4.070   &2.813   &4.019   &26	\\
BR J0419-5716     &4.461   &2.820   &4.406   &26	\\
Q 0420+007        &2.918   &1.673   &2.879   &21   	\\ 
Q 0420-3851       &3.1230  &2.094   &3.082   &21   	\\ 
BR J0426-2202     &4.320   &2.544   &4.267   &26	\\
Q 0428-1342       &3.244   &1.965   &3.200   &21   	\\ 
Q 0454-220        &0.5340  &0.1199  &0.5187  &3    	\\ 
Q 0454+039        &1.3450  &0.9672  &1.3216  &3    	\\ 
Q 0457+024        &2.38    &1.645   &2.346   &16   	\\ 
MG 0504+0303      &2.46    &1.803   &2.425   &2    	\\ 
Q 0521-365        &0.0566  &0.0080  &0.0460  &3    	\\ 
PMN J0525-3343    &4.383   &2.829   &4.329   &26	\\
BR J0529-3526     &4.413   &3.023   &4.359   &26	\\
BR J0529-3552     &4.172   &2.821   &4.120   &26	\\
Q 0537-441        &0.8940  &0.5139  &0.6300  &3    	\\ 
Q 0548-322        &0.0690  &0.0080  &0.0583  &3    	\\ 
Q 0552+398        &2.36    &1.644   &2.325   &16   	\\ 
Q 0558-504        &0.1370  &0.0080  &0.1256  &3    	\\ 
Q 0624+691        &0.3700  &0.0080  &0.3563  &3    	\\ 
\hline
\end{tabular}
\end{table}

\begin{table}
\centering
\setcounter{table}{0}
\caption{{\it continued}}  
\begin{tabular}{lllll}
\hline
Quasar	&$z_{\rm em}$	&$z_{\rm min}$	&$z_{\rm max}$	&Ref$^a$\\
\hline
Q 0636+6801       &3.178   &2.019   &3.132   &21   	\\ 
Q 0637-752        &0.6560  &0.0080  &0.6251  &3    	\\ 
Q 0642+4454       &3.408   &2.192   &3.362   &21   	\\ 
Q 0702+646        &0.0795  &0.0080  &0.0687  &3    	\\ 
BR J0714-6455     &4.462   &3.050   &4.407   &26	\\
Q 0731+6519       &3.038   &2.019   &2.993   &21   	\\ 
Q 0735+178        &0.4240  &0.0765  &0.4098  &3    	\\ 
Q 0736+017        &0.1910  &0.0080  &0.1791  &3    	\\ 
Q 0742+318        &0.4620  &0.0080  &0.4474  &3    	\\ 
Q 0743-673        &1.5130  &1.0302  &1.4879  &3    	\\ 
GB 0749+4239      &3.59    &2.185   &3.544   &2    	\\ 
PC 0751+5623      &4.281   &3.526   &4.228   &1    	\\ 
Q 0754+100        &0.6700  &0.0080  &0.6257  &3    	\\ 
Q 0754+394        &0.0958  &0.0080  &0.0848  &3    	\\ 
Q 0804+761        &0.1000  &0.0080  &0.0890  &3    	\\ 
Q 0805+0441       &2.880   &1.838   &2.834   &21   	\\ 
Q 0812+332        &2.42    &1.677   &2.385   &16   	\\ 
Q 0819-032        &2.35    &1.704   &2.319   &16   	\\ 
Q 0820+296        &2.37    &1.644   &2.333   &16   	\\ 
MG 0830+1009      &3.75    &2.040   &3.703   &2    	\\ 
Q 0830+1133       &2.979   &1.797   &2.936   &21   	\\ 
Q 0831+1238       &2.748   &1.961   &2.706   &21   	\\ 
Q 0837-120        &0.1980  &0.0080  &0.1860  &3    	\\ 
Q 0844+349        &0.0640  &0.0080  &0.0534  &3    	\\ 
Q 0846+152        &2.64    &1.831   &2.599   &16   	\\ 
MG 0848+1533      &2.01    &1.735   &1.980   &2    	\\ 
Q 0849+080        &0.0620  &0.0080  &0.0514  &3    	\\ 
Q 0851+202        &0.3060  &0.0080  &0.2929  &3    	\\ 
Q 0855+182        &2.62    &1.682   &2.580   &16   	\\ 
Q 0903+155        &2.68    &1.659   &2.645   &16   	\\ 
MG 0906+0406      &3.20    &1.811   &3.158   &2    	\\ 
Q 0906+484        &0.1180  &0.0080  &0.1068  &3    	\\ 
Q 0910+403        &0.9360  &0.0080  &0.9166  &3    	\\ 
Q 0914-621        &0.0573  &0.0080  &0.0467  &3    	\\ 
Q 0916+555        &0.1235  &0.0080  &0.1123  &3    	\\ 
Q 0932+3646       &2.84    &1.634   &2.814   &21   	\\ 
Q 0933+733        &2.53    &1.651   &2.493   &16   	\\ 
Q 0938+1159       &3.19    &1.634   &3.149   &21   	\\ 
Q 0941+2608       &2.913   &1.731   &2.867   &21   	\\ 
Q 0953+414        &0.2390  &0.0080  &0.2266  &3    	\\ 
Q 0955+326        &0.5330  &0.0080  &0.5177  &3    	\\ 
Q 0956+1217       &3.306   &2.159   &3.263   &21   	\\ 
Q 0957+561        &1.4050  &0.8179  &1.3810  &3    	\\ 
Q 0958+551        &1.7324  &1.1762  &1.4513  &3    	\\ 
Q 1001+291        &0.3290  &0.0080  &0.3157  &3    	\\ 
Q 1004+130        &0.2410  &0.0080  &0.2286  &3    	\\ 
Q 1004+1411       &2.707   &1.786   &2.672   &21   	\\ 
Q 1007+417        &0.6110  &0.0080  &0.5949  &3    	\\ 
Q 1009-0252       &2.75    &1.651   &2.708   &4    	\\
Q 1011-282        &0.6110  &0.0080  &0.1310  &3    	\\ 
Q 1011-0144       &2.24    &1.669   &2.204   &4    	\\
Q 1011+250        &1.6310  &0.9718  &1.5500  &3    	\\ 
Q 1012+008        &0.1850  &0.0080  &0.1732  &3    	\\ 
Q 1012-0206       &2.14    &1.634   &2.104   &4    	\\
GB 1013+2052      &3.11    &1.945   &3.069   &2    	\\ 
Q 1014+0023       &2.29    &1.634   &2.591   &4    	\\
Q 1016-0039       &2.18    &1.649   &2.144   &4    	\\
Q 1017+1055       &3.158   &2.114   &3.127   &21   	\\ 
Q 1017+280        &1.9280  &0.9971  &1.4678  &3    	\\ 
Q 1018-0005       &2.60    &1.789   &2.560   &4    	\\
Q 1020+0028       &1.90    &1.680   &1.872   &4    	\\
Q 1021-0037       &2.547   &1.887   &2.513   &21   	\\ 
Q 1024+0030       &2.17    &1.717   &2.135   &4    	\\
Q 1025-0030       &2.87    &1.885   &2.833   &4    	\\
\hline
\end{tabular}
\end{table}

\begin{table}
\centering
\setcounter{table}{0}
\caption{{\it continued}}  
\begin{tabular}{lllll}
\hline
Quasar	&$z_{\rm em}$	&$z_{\rm min}$	&$z_{\rm max}$	&Ref$^a$\\
\hline
RX J1028-0844     &4.276   &2.533   &4.223   &26	\\
Q 1028+313        &0.1770  &0.0080  &0.1652  &3    	\\ 
Q 1029-140        &0.0860  &0.0080  &0.0751  &3    	\\ 
Q 1033+1342       &3.07    &1.800   &3.048   &21   	\\ 
BR B1033-0327     &4.509   &2.91    &4.454   &22,23	\\ 
Q 1038+528        &2.30    &1.677   &2.262   &16   	\\ 
GB 1041+3014      &2.99    &1.735   &2.950   &2    	\\ 
Q 1047+550        &2.1650  &1.3299  &1.5159  &3    	\\ 
BRI B1050-0000    &4.286   &2.83    &4.233   &22   	\\ 
Q 1100+772        &0.3110  &0.0080  &0.2979  &3    	\\ 
Q 1100-264        &2.1450  &1.1551  &1.5500  &3    	\\ 
MG 1101+0248      &2.51    &1.736   &2.475   &2    	\\ 
Q 1103-006        &0.4260  &0.0080  &0.4117  &3    	\\ 
BRI B1110+0106    &3.918   &2.58    &3.869   &22   	\\ 
Q 1115+080        &1.7180  &0.4066  &0.6330  &3    	\\ 
Q 1115+080        &1.7180  &0.9595  &1.5500  &3    	\\ 
Q 1116+215        &0.1770  &0.0080  &0.1652  &3    	\\ 
Q 1123+264        &2.35    &1.645   &2.317   &16   	\\ 
Q 1124+5706       &2.890   &1.762   &2.851   &21   	\\ 
Q 1127+078        &2.66    &1.644   &2.621   &16   	\\ 
Q 1128+105        &2.65    &2.040   &2.610   &16   	\\ 
Q 1131-0043       &2.16    &1.653   &2.128   &4    	\\
Q 1132-0054       &2.76    &1.717   &2.718   &4    	\\
Q 1135-0255       &2.41    &1.739   &2.373   &4    	\\
Q 1136-135        &0.5570  &0.0080  &0.5414  &3    	\\ 
Q 1136+122        &2.90    &1.781   &2.862   &16   	\\ 
Q 1137+660        &0.6460  &0.0080  &0.6295  &3    	\\ 
Q 1138-0107       &2.76    &1.953   &2.718   &4    	\\
Q 1139-0139       &1.93    &1.634   &1.884   &4    	\\
Q 1139-0037       &1.91    &1.634   &1.896   &4    	\\
Q 1142+0138       &2.42    &1.791   &2.390   &4    	\\
Q 1142+1015       &3.152   &2.127   &3.109   &21   	\\ 
Q 1143+0142       &2.28    &1.634   &2.248   &4    	\\
Q 1143+099        &2.60    &1.676   &2.567   &16   	\\ 
Q 1144+115        &2.51    &1.682   &2.471   &16   	\\ 
Q 1144+0140       &2.59    &1.667   &2.551   &4    	\\
Q 1145-0039       &1.94    &1.634   &1.912   &4    	\\
Q 1145+0121       &2.08    &1.721   &2.045   &4    	\\
Q 1146+0207       &2.06    &1.634   &2.025   &4    	\\
Q 1147+084        &2.61    &1.854   &2.577   &16   	\\ 
GB 1147+4348      &3.02    &2.035   &2.980   &2    	\\ 
Q 1148-0007       &1.977   &1.634   &1.947   &4    	\\
Q 1148+0055       &1.89    &1.667   &1.858   &4    	\\
Q 1148+549        &0.9690  &0.0080  &0.9493  &3    	\\ 
Q 1151+117        &0.1760  &0.0080  &0.1642  &3    	\\ 
Q 1156+295        &0.7290  &0.0080  &0.7117  &3    	\\ 
Q 1159+0039       &2.586   &1.671   &2.550   &21   	\\ 
Q 1202+281        &0.1650  &0.0080  &0.1534  &3    	\\ 
Q 1205-3014       &3.036   &2.045   &2.996   &21   	\\ 
Q 1206+1155       &3.106   &2.039   &3.073   &21   	\\ 
Q 1206+1500       &2.60    &1.793   &2.568   &4    	\\
Q 1206+1727       &2.36    &1.634   &2.321   &4    	\\
Q 1206+459        &1.1580  &0.4231  &0.6300  &3    	\\ 
Q 1206+459        &1.1580  &0.8426  &1.1364  &3    	\\ 
Q 1209+1046       &2.20    &1.634   &2.163   &4    	\\
Q 1209+1524       &3.06    &1.634   &3.021   &4    	\\
Q 1211+143        &0.0850  &0.0080  &0.0742  &3    	\\ 
Q 1212+1551       &1.95    &1.665   &1.918   &4    	\\
Q 1212+1045       &1.95    &1.634   &1.922   &4    	\\
Q 1212+0854       &2.35    &1.634   &2.319   &4    	\\
Q 1213+1015       &2.52    &1.634   &2.482   &4    	\\
Q 1213+0922       &2.72    &1.675   &2.681   &4    	\\
Q 1215+1244       &2.08    &1.634   &2.048   &4    	\\
Q 1215+1202       &2.83    &1.634   &2.788   &4    	\\
\hline
\end{tabular}
\end{table}

\begin{table}
\centering
\setcounter{table}{0}
\caption{{\it continued}}  
\begin{tabular}{lllll}
\hline
Quasar	&$z_{\rm em}$	&$z_{\rm min}$	&$z_{\rm max}$	&Ref$^a$\\
\hline
Q 1215+303        &0.2370  &0.0080  &0.2246  &3    	\\ 
Q 1216+069        &0.3340  &0.0080  &0.3207  &3    	\\ 
Q 1216+1517       &1.83    &1.723   &1.802   &4    	\\
Q 1216+1754       &1.81    &1.634   &1.781   &4    	\\
Q 1216+1656       &2.83    &1.659   &2.791   &4    	\\
Q 1216+0947       &2.31    &1.645   &2.279   &4    	\\
Q 1217+023        &0.2400  &0.0080  &0.2276  &3    	\\ 
Q 1218+304        &0.1300  &0.0080  &0.1187  &3    	\\ 
Q 1219+755        &0.0700  &0.0080  &0.0593  &3    	\\ 
Q 1219+1140       &2.18    &1.634   &2.147   &4    	\\
Q 1222+228        &2.0400  &0.4647  &0.6316  &3    	\\ 
Q 1222+1053       &2.30    &1.641   &2.263   &4    	\\
Q 1223+1059       &2.32    &1.643   &2.288   &4    	\\
Q 1223+1723       &2.42    &1.659   &2.386   &4    	\\
Q 1224+1244       &2.14    &1.634   &2.110   &4    	\\
Q 1225+1512       &2.01    &1.797   &1.977   &4    	\\
Q 1225+1610       &2.23    &1.663   &2.200   &4    	\\
Q 1225+317        &2.2190  &1.1263  &1.5500  &3    	\\ 
Q 1226+1035       &2.32    &1.634   &2.287   &4    	\\
Q 1226+1115       &1.98    &1.634   &1.950   &4    	\\
Q 1226+1639       &2.25    &1.634   &2.216   &4    	\\
Q 1226+023        &0.1580  &0.0080  &0.1464  &3    	\\ 
Q 1227+1215       &2.17    &1.624   &2.138   &4    	\\
Q 1228+1808       &2.64    &1.780   &2.607   &4    	\\
Q 1228+077        &2.39    &1.691   &2.354   &16   	\\ 
Q 1229+1414       &2.90    &1.764   &2.862   &4    	\\
Q 1229+1531       &2.27    &1.634   &2.237   &4    	\\
Q 1229-021        &1.0380  &0.4738  &0.6320  &3    	\\ 
Q 1229+204        &0.0640  &0.0080  &0.0534  &3    	\\ 
Q 1230+1042       &2.43    &1.634   &2.396   &4    	\\
Q 1230+1318       &2.29    &1.634   &2.257   &4    	\\
Q 1230+1627B      &2.70    &1.634   &2.663   &4    	\\
Q 1230+0941       &1.84    &1.641   &1.812   &4    	\\
Q 1232-0051       &2.78    &1.782   &2.745   &4    	\\
Q 1232+1139       &2.87    &1.848   &2.831   &4    	\\
Q 1234+0122       &2.03    &1.634   &1.996   &4    	\\
Q 1235+1807A      &2.41    &1.782   &2.371   &4    	\\
Q 1236-0043       &1.84    &1.690   &1.815   &4    	\\
Q 1236-0207       &2.25    &1.729   &2.213   &4    	\\
Q 1237+1515       &2.04    &1.634   &2.009   &4    	\\
Q 1237+0107       &1.81    &1.733   &1.780   &4    	\\
Q 1237+1508       &2.07    &1.634   &2.035   &4    	\\
Q 1237+1212       &2.31    &1.634   &2.281   &4    	\\
Q 1239+1435       &1.93    &1.634   &1.900   &4    	\\
Q 1239+0249       &2.22    &1.719   &2.184   &4    	\\
Q 1240+1504       &1.85    &1.634   &1.823   &4    	\\
Q 1241+176        &1.2730  &0.4066  &0.6320  &3    	\\ 
Q 1241+176        &1.2730  &0.7657  &1.2503  &3    	\\ 
Q 1242+0213       &1.99    &1.634   &1.958   &4    	\\
Q 1242+0006       &2.08    &1.634   &2.045   &4    	\\
Q 1242+1732       &1.83    &1.696   &1.805   &4    	\\
Q 1242+1737       &1.86    &1.634   &1.828   &4    	\\
Q 1244+1129       &3.16    &2.101   &3.118   &4    	\\
Q 1244+1642       &2.87    &1.848   &2.826   &4    	\\
Q 1246-0059       &2.45    &1.669   &2.415   &4    	\\
Q 1246+0032       &2.31    &1.651   &2.273   &4    	\\
Q 1247+267        &2.0380  &0.9211  &1.5500  &3    	\\ 
Q 1248+401        &1.0300  &0.3984  &0.6028  &3    	\\ 
Q 1248+401        &1.0300  &0.8919  &1.0097  &3    	\\ 
Q 1253-055        &0.5380  &0.0080  &0.5226  &3    	\\ 
Q 1259+593        &0.4720  &0.0080  &0.4573  &3    	\\ 
Q 1302-102        &0.2860  &0.0080  &0.2731  &3    	\\ 
Q 1307+085        &0.1550  &0.0080  &0.1435  &3    	\\ 
Q 1308+326        &0.9960  &0.4670  &0.6310  &3    	\\ 
\hline
\end{tabular}
\end{table}

\begin{table}
\centering
\setcounter{table}{0}
\caption{{\it continued}}  
\begin{tabular}{lllll}
\hline
Quasar	&$z_{\rm em}$	&$z_{\rm min}$	&$z_{\rm max}$	&Ref$^a$\\
\hline
Q 1308-0214       &2.85    &1.892   &2.811   &4    	\\
Q 1308-0104       &2.59    &1.634   &2.549   &4    	\\
Q 1309+355        &0.1840  &0.0080  &0.1722  &3    	\\ 
BR J1310-1740     &4.185   &2.508   &4.133   &26	\\
Q 1312+043        &2.35    &1.813   &2.319   &16   	\\ 
Q 1313+0107       &2.39    &1.647   &2.359   &4    	\\
PSS J1317+3531    &4.365   &2.978   &4.311   &1    	\\
Q 1317+277        &1.0220  &0.2503  &1.0018  &3    	\\ 
Q 1318+290B       &0.5490  &0.3757  &0.5335  &3    	\\ 
Q 1318-0150       &2.01    &1.651   &1.980   &4    	\\
Q 1318-113        &2.3080  &1.896   &2.273   &16   	\\ 
Q 1320+0048       &1.96    &1.655   &1.925   &4    	\\
Q 1323-0248       &2.12    &1.661   &2.090   &4    	\\
Q 1324-0212       &1.89    &1.634   &1.857   &4    	\\
Q 1327-206        &1.1690  &1.1243  &1.1473  &3    	\\ 
Q 1328+0223       &2.15    &1.937   &2.122   &4    	\\
BRI B1328-0433    &4.217   &2.24    &4.165   &22   	\\ 
Q 1329+0231       &2.43    &1.663   &2.400   &4    	\\
Q 1329+0018       &2.35    &1.661   &2.318   &4    	\\
Q 1329+4117       &1.9350  &0.4853  &0.6318  &3    	\\    
Q 1331+170        &2.0840  &1.2621  &1.5500  &3    	\\ 
Q 1333+176        &0.5540  &0.3902  &0.5385  &3    	\\ 
Q 1334+246        &0.1070  &0.0080  &0.0959  &3    	\\ 
Q 1334-0033       &2.78    &1.634   &2.745   &4    	\\
Q 1334+0212       &2.38    &1.634   &2.350   &4    	\\
BRI B1335-0417    &4.396   &3.08    &4.342   &22   	\\ 
Q 1336+0210       &1.96    &1.634   &1.932   &4    	\\
GB 1338+3809      &3.10    &1.737   &3.059   &2    	\\ 
Q 1338+101        &2.45    &1.724   &2.412   &16   	\\ 
Q 1338+416        &1.2190  &0.4066  &0.6324  &3    	\\ 
Q 1338+416        &1.2190  &0.8684  &1.1968  &3    	\\ 
Q 1340+0959       &2.942   &1.894   &2.897   &21   	\\ 
Q 1344+0137       &1.92    &1.634   &1.886   &4    	\\
Q 1345-0137       &1.93    &1.634   &1.900   &4    	\\
Q 1345-0120       &2.95    &1.926   &2.906   &4    	\\
Q 1346+0121A      &1.93    &1.634   &1.901   &4    	\\
Q 1346-036        &2.36    &1.653   &2.327   &16   	\\ 
Q 1351+640        &0.0880  &0.0080  &0.0771  &3    	\\ 
Q 1352+183        &0.1520  &0.0080  &0.1405  &3    	\\ 
Q 1352+108        &3.18    &1.928   &3.137   &16   	\\ 
Q 1353+186        &0.0505  &0.0080  &0.0400  &3    	\\ 
Q 1354+195        &0.7200  &0.3593  &0.6330  &3    	\\ 
Q 1355-416        &0.3130  &0.0080  &0.2999  &3    	\\ 
Q 1356+581        &1.3710  &0.5218  &0.6310  &3    	\\ 
Q 1358+115        &2.59    &1.677   &2.550   &16   	\\ 
Q 1358+3908       &3.3     &2.221   &3.237   &21   	\\ 
Q 1400+0935       &2.980   &2.022   &2.930   &21   	\\ 
Q 1402-012        &2.52    &1.789   &2.479   &16   	\\ 
Q 1402+044        &3.20    &2.340   &3.160   &16   	\\ 
Q 1406+123        &2.94    &2.018   &2.903   &16   	\\ 
Q 1407+265        &0.9440  &0.0080  &0.9246  &3    	\\ 
Q 1410+096        &3.21    &2.099   &3.169   &16   	\\ 
FIRST J1410+3409  &4.351   &3.026   &3.578   &26	\\    
                  &        &3.602   &4.297   &26	\\
Q 1411+442        &0.0900  &0.0080  &0.0791  &3    	\\ 
GB 1413+3720      &2.36    &1.735   &2.326   &2    	\\ 
Q 1415+451        &0.1140  &0.0080  &0.1029  &3    	\\ 
Q 1416-129        &0.1290  &0.0080  &0.1177  &3    	\\ 
Q 1418+546        &0.1520  &0.0080  &0.1405  &3    	\\ 
Q 1419+480        &0.0720  &0.0080  &0.0613  &3    	\\ 
Q 1421+330        &1.9040  &1.0311  &1.5500  &3    	\\ 
Q 1425+267        &0.3620  &0.2409  &0.3484  &3    	\\ 
Q 1426+015        &0.0860  &0.0080  &0.0751  &3    	\\ 
Q 1428+0202       &2.11    &1.634   &2.075   &4    	\\
\hline
\end{tabular}
\end{table}

\begin{table}
\centering
\setcounter{table}{0}
\caption{{\it continued}}  
\begin{tabular}{lllll}
\hline
Quasar	&$z_{\rm em}$	&$z_{\rm min}$	&$z_{\rm max}$	&Ref$^a$\\
\hline
Q 1429-0053       &2.08    &1.719   &2.047   &4    	\\
Q 1429+118        &3.00    &1.958   &2.963   &16   	\\ 
PSS J1430+2828    &4.306   &2.777   &4.253   &1    	\\
Q 1433+0223       &2.14    &1.634   &2.111   &4    	\\
Q 1433-0025       &2.04    &1.634   &2.012   &4    	\\
PSS J1435+3057    &4.297   &2.905   &4.244   &1    	\\
GB 1436+4431      &2.10    &1.769   &2.069   &2    	\\ 
Q 1439+0047       &1.86    &1.649   &1.828   &4    	\\
Q 1440-0024       &1.81    &1.634   &1.786   &4    	\\
Q 1440+356        &0.0781  &0.0080  &0.0673  &3    	\\ 
Q 1444+407        &0.2670  &0.0080  &0.2543  &3    	\\ 
Q 1444+0126       &2.21    &1.717   &2.174   &4    	\\
Q 1444-0112       &2.15    &1.651   &2.121   &4    	\\
Q 1451-375        &0.3140  &0.0080  &0.3009  &3    	\\ 
Q 1455+123        &3.08    &1.830   &3.033   &16   	\\ 
PSS J1456+2007    &4.249   &2.878   &4.197   &26	\\
MG 1500+0431      &3.67    &2.606   &3.623   &1    	\\
Q 1503+118        &2.78    &1.957   &2.740   &16   	\\ 
GB 1508+5714      &4.283   &2.73    &4.230   &22   	\\ 
Q 1512+370        &0.3710  &0.0080  &0.3573  &3    	\\ 
MG 1519+1806      &3.06    &1.955   &3.019   &2    	\\ 
GB 1520+4347      &2.18    &1.775   &2.148   &2    	\\ 
Q 1522+101        &1.3210  &0.0080  &0.6310  &3    	\\ 
Q 1522+101        &1.3210  &0.8803  &1.2978  &3    	\\ 
Q 1525+227        &0.2530  &0.0080  &0.2405  &3    	\\ 
GB 1526+6701      &3.02    &1.955   &2.980   &2    	\\ 
Q 1526+285        &0.4500  &0.0080  &0.2428  &3    	\\ 
Q 1538+477        &0.7700  &0.3326  &0.6326  &3    	\\ 
Q 1545+210        &0.2640  &0.0080  &0.2514  &3    	\\ 
Q 1548+0917       &2.749   &1.874   &2.707   &21   	\\ 
PC 1548+4637      &3.544   &2.607   &3.499   &1    	\\
Q 1553+113        &0.3600  &0.0080  &0.3464  &3    	\\ 
Q 1556+273        &0.0899  &0.0080  &0.0790  &3    	\\ 
MG 1557+0313      &3.891   &2.66    &3.842   &22   	\\ 
MG 1559+1405      &2.24    &1.737   &3.059   &2    	\\ 
Q 1600+0729       &4.38    &3.062   &4.326   &1    	\\ 
BR J1603+0721     &4.385   &3.062   &4.331   &26	\\
Q 1607+1819       &3.123   &1.814   &3.0918  &21   	\\ 
Q 1612+261        &0.1310  &0.0080  &0.1197  &3    	\\ 
Q 1613+658        &0.1290  &0.0080  &0.1177  &3    	\\ 
Q 1623+268A       &2.47    &1.644   &2.433   &16   	\\ 
Q 1623+268B       &2.54    &1.644   &2.502   &16   	\\ 
Q 1630+377        &1.4710  &0.0080  &0.6320  &3    	\\ 
Q 1630+377        &1.4710  &0.8641  &1.4463  &3    	\\ 
Q 1631+3722       &2.940   &1.785   &2.906   &21   	\\ 
PSS J1633+1411    &4.351   &2.536   &4.297   &26	\\
Q 1634+706        &1.3340  &0.5547  &1.3107  &3    	\\ 
Q 1641+399        &0.5950  &0.0080  &0.5791  &3    	\\ 
PC 1640+4628      &3.700   &2.604   &3.653   &1    	\\
PSS J1646+5514    &4.037   &2.772   &3.987   &26	\\
Q 1704+608        &0.3710  &0.0080  &0.3573  &3    	\\ 
Q 1705+0152       &2.576   &1.669   &2.537   &21   	\\ 
Q 1715+535        &1.9290  &1.1009  &1.5500  &3    	\\ 
Q 1718+481        &1.0840  &0.0080  &1.0632  &3    	\\ 
PSS J1721+3256    &4.031   &2.791   &3.981   &26	\\
Q 1721+343        &0.2060  &0.0080  &0.1939  &3    	\\ 
Q 1726+3425       &2.429   &1.669   &2.393   &21   	\\ 
Q 1727+502        &0.0550  &0.0080  &0.0445  &3    	\\ 
Q 1738+3502       &3.240   &2.093   &3.197   &21   	\\ 
GB 1745+6227      &3.901   &2.47    &3.852   &22   	\\ 
Q 1803+676        &0.1360  &0.0080  &0.1246  &3    	\\ 
Q 1807+698        &0.0512  &0.0080  &0.0407  &3    	\\ 
Q 1821+643        &0.2970  &0.0080  &0.2840  &3    	\\ 
Q 1831+731        &0.1230  &0.0080  &0.1118  &3    	\\ 
\hline
\end{tabular}
\end{table}

\begin{table}
\centering
\setcounter{table}{0}
\caption{{\it continued}}  
\begin{tabular}{lllll}
\hline
Quasar	&$z_{\rm em}$	&$z_{\rm min}$	&$z_{\rm max}$	&Ref$^a$\\
\hline
Q 1833+326        &0.0590  &0.0080  &0.0484  &3    	\\ 
Q 1836+5108       &2.827   &1.920   &2.789   &21   	\\ 
Q 1839-785        &0.0743  &0.0080  &0.0636  &3    	\\ 
Q 1845+797        &0.0556  &0.0080  &0.0450  &3    	\\ 
Q 1912-550        &0.4020  &0.1769  &0.2041  &3    	\\ 
Q 1928+738        &0.3020  &0.0080  &0.2890  &3    	\\ 
PKS 1937-101      &3.787   &2.442   &3.739   &21   	\\ 
Q 2000-3300       &3.783   &2.521   &3.729   &21   	\\ 
Q 2005-489        &0.0710  &0.0080  &0.0603  &3    	\\ 
Q 2038-0116       &2.783   &1.887   &2.745   &21   	\\ 
Q 2045-377        &1.8000  &1.0040  &1.5500  &3    	\\ 
Q 2048+3116       &3.198   &1.830   &3.143   &21   	\\ 
Q 2050-359        &3.49    &2.605   &3.445   &1    	\\
Q 2112+0555       &0.4660  &0.1105  &0.4513  &3,24,25 	\\
Q 2113-4345       &2.05    &1.664   &2.023   &4    	\\
Q 2113-4534       &2.54    &1.969   &2.506   &4    	\\
Q 2114-4346       &2.04    &1.606   &2.011   &4    	\\
Q 2115-4434       &2.16    &1.755   &2.128   &4    	\\
Q 2117-4703       &2.26    &1.849   &2.223   &4    	\\
Q 2122-4231       &2.27    &1.550   &2.233   &4    	\\
Q 2126-1551       &3.2660  &2.011   &3.218   &21   	\\ 
Q 2126-4618       &1.89    &1.715   &1.859   &4    	\\
Q 2127-4528       &2.71    &2.018   &2.676   &4    	\\
Q 2128-123        &0.5010  &0.0940  &0.4860  &3    	\\ 
Q 2130+099        &0.0610  &0.0080  &0.0504  &3    	\\ 
Q 2131-4257       &2.10    &1.590   &2.065   &4    	\\
PMN J2134-0419    &4.334   &2.903   &4.281   &26	\\
Q 2134-4239       &1.80    &1.590   &1.776   &4    	\\
Q 2134-147        &0.2000  &0.0080  &0.1880  &3    	\\ 
Q 2135-4632       &2.21    &1.879   &2.182   &4    	\\
Q 2136+141        &2.43    &1.784   &2.390   &16   	\\ 
Q 2139-4434       &3.23    &2.373   &3.188   &4    	\\
Q 2141+175        &0.2130  &0.0080  &0.2009  &3    	\\ 
Q 2145+067        &0.9900  &0.9426  &0.9701  &3    	\\ 
MG 2152+1420      &2.56    &1.800   &2.524   &2    	\\ 
Q 2153-2056       &1.85    &1.634   &1.821   &4    	\\
Q 2155-304        &0.1170  &0.0080  &0.1058  &3    	\\ 
Q 2159-2058       &2.12    &1.634   &2.089   &4    	\\
Q 2201+315        &0.2970  &0.0080  &0.2840  &3    	\\ 
Q 2203-2145       &2.27    &1.692   &2.240   &4    	\\
Q 2203-1833       &2.73    &1.849   &2.691   &4    	\\
Q 2205-2014       &2.64    &1.652   &2.599   &4    	\\
MG 2206+1753      &3.14    &1.769   &3.099   &2    	\\ 
Q 2209-1842       &2.09    &1.634   &2.061   &4    	\\
Q 2209+184        &0.0700  &0.0080  &0.0593  &3    	\\ 
Q 2211-1915       &1.95    &1.634   &1.923   &4    	\\
BR B2212-1626     &3.990   &2.69    &3.940   &22   	\\ 
Q 2214+139        &0.0658  &0.0080  &0.0551  &3    	\\ 
BR J2216-6714     &4.469   &2.795   &4.414   &26	\\
MG 2222+0511      &2.32    &1.800   &2.287   &2    	\\ 
GB 2223+2024      &3.56    &2.101   &3.514   &2    	\\ 
Q 2231+0125       &1.90    &1.634   &1.871   &4    	\\
Q 2231-0212       &1.90    &1.634   &1.871   &4    	\\
Q 2233+1341       &3.209   &2.216   &3.167   &21   	\\ 
Q 2233+1310       &3.298   &2.134   &3.252   &21   	\\ 
Q 2241+0014       &2.14    &1.657   &2.099   &4    	\\
Q 2243+0141       &2.30    &1.663   &2.267   &4    	\\
Q 2244-0234       &1.97    &1.787   &1.940   &4    	\\
Q 2244-0105       &2.04    &1.634   &2.010   &4    	\\
Q 2246-0006       &2.05    &1.651   &2.019   &4    	\\
BR B2248-1242     &4.161   &2.94    &4.109   &22   	\\ 
MG 2251+2429      &2.33    &2.019   &2.297   &2    	\\ 
Q 2251-178        &0.0680  &0.0080  &0.0573  &3    	\\ 
Q 2251+113        &0.3230  &0.1310  &0.3098  &3    	\\ 
\hline
\end{tabular}
\end{table}

\begin{table}
\centering
\setcounter{table}{0}
\caption{{\it continued}}  
\begin{tabular}{lllll}
\hline
Quasar	&$z_{\rm em}$	&$z_{\rm min}$	&$z_{\rm max}$	&Ref$^a$\\
\hline
MG 2254+0227      &2.09    &2.767   &2.059   &2    	\\ 
Q 2256+017        &2.67    &1.786   &2.629   &16   	\\ 
Q 2302+029        &1.0440  &0.3942  &0.6290  &3    	\\ 
Q 2302+029        &1.0440  &0.8060  &1.0236  &3    	\\ 
Q 2308+098        &0.4320  &0.0080  &0.4177  &3    	\\ 
Q 2311-0341       &3.048   &1.714   &3.001   &21   	\\ 
MG 2320+0755      &2.09    &1.780   &2.059   &2    	\\ 
Q 2326-477        &1.2990  &0.9164  &1.2760  &3    	\\ 
BR J2328-4513     &4.359   &2.926   &4.305   &26	\\
PC $2331+0216$    &4.093   &3.115   &4.042   &1    	\\  
Q 2334+1041       &2.243   &1.634   &2.211   &21   	\\ 
Q 2344+092        &0.6720  &0.0080  &0.6288  &3    	\\ 
BR J2349-3712     &4.208   &2.847   &4.156   &26	\\
Q 2351+1042       &2.379   &1.632   &2.345   &21   	\\ 
Q 2351+0120       &2.07    &1.634   &2.039   &4    	\\
Q 2351-1154       &2.67    &1.632   &2.633   &21   	\\ 
Q 2352+0205       &2.19    &1.634   &2.158   &4    	\\
Q 2354-0134       &2.21    &1.665   &2.178   &4    	\\
Q 2356+0139       &2.07    &1.661   &2.039   &4    	\\
Q 2356+0237       &2.50    &1.634   &2.465   &4    	\\
Q 2359+0653       &3.238   &1.632   &3.203   &21   	\\ 
Q 2359+0023       &2.897   &1.714   &2.857   &21   	\\
\hline
\end{tabular}
\vspace{0.5cm}
\begin{small}
$^a$References: 1) Storrie-Lombardi \& Wolfe 2000; 2) Storrie-Lombardi
\& Hook 2000; 3) Lanzetta, Wolfe, \& Turnshek 1995; 4) Wolfe et
al. 1995; 5) Sargent, Steidel, \& Boksenberg 1989; 6) Turnshek et
al. 1989; 7) Wolfe et al. 1993; 8) Lu et al. 1993; 9) Lu \& Wolfe; 10)
Virgilio et al. 1995; 11) Pettini et al. 1994; 12) Francis \& Hewett
1993; 13) Savaglio et al. 1994; 14) Sargent, Boksenberg, \& Steidel
1988; 15) Black, Chaffee, \& Foltz 1987; 16) Wolfe et al. 1986; 17)
Wolfe et al. 1994; 18) Rauch et al. 1990; 19) Williger et al. 1989;
20) Meyer, Lanzetta \& Wolfe 1995; 21) Lanzetta et al. 1991; 22)
Storrie-Lombardi, McMahon, Irwin, \& Hazard 1996; 23)
Storrie-Lombardi, Irwin \& McMahon 1996; 24) Jannuzi et al. 1998; 25)
Fynbo, Moller \& Thomsen 2001; 26) P\'eroux et al. 2001 
\end{small}
\end{table}

\bsp

\label{lastpage}

\end{document}